\documentclass[twocolumn,preprintnumbers,amsmath,amssymb,prb,showpacs]{revtex4}
\usepackage{graphicx}
\usepackage{dcolumn}
\usepackage{bm}

\begin{document}
\title{Doping graphene by adsorption of polar molecules 
at the oxidized zigzag edges}
\author{Julia Berashevich and Tapash Chakraborty}
\email{chakrabt@cc.umanitoba.ca}
\affiliation{Department of Physics and Astronomy, University of 
Manitoba, Winnipeg, Canada, R3T 2N2}

\begin{abstract}
We have theoretically investigated the electronic and magnetic properties of 
graphene whose zigzag edges are oxidized. 
The alteration of these properties 
by adsorption of $\mathrm{H_{2}O}$ and $\mathrm{NH_3}$ molecules
have been considered. It was found that the adsorbed molecules 
form a cluster along the oxidized zigzag edges of graphene
due to interaction with the electro-negative oxygen. 
Graphene tends to donate a charge to the adsorbates through the oxygen atoms 
and the efficiency of donation depends on the intermolecular distance 
and on the location of the adsorbed molecules relative to the plane of graphene. 
It was found that by appropriate selection of the adsorbates, a controllable and gradual growth 
of $p$-doping in graphene with a variety of adsorbed molecules can be achieved.
\end{abstract}
\maketitle

\section{Introduction}
The edges of graphene, in particular the zigzag edges, 
play the most important role in defining the electronic properties of graphene \cite{nakada1996}.
Geometry of the zigzag edges localizes the electrons with
maximum of the electron density at the border carbon atoms.
The localized states form the flat conduction and valence bands near the Dirac points
and the size of the gap between them is defined by the 
contribution of the confinement effect and shape of the edges \cite{conf,barone}.
Any changes in the edge geometry, 
such as termination, reconstruction or distortion, 
would modify the electronic properties of graphene.
Termination of the edges most often is realized through 
the edge hydrogenation, which alters the $sp^2$ network at the zigzag edges \cite{barone}.
For the pure graphene edges, there is a mixture of the $sp$ and $sp^2$ hybridization, 
while for hydrogenated graphene the $sp^2$ and $sp^3$ hybridization dominates 
providing an increase in the size of the gap.
Additionally to hydrogenation, structural changes in the zigzag edges, 
such as reconstruction and aromaticity, 
also have effect on the electronic properties of graphene \cite{wassmann2008}. 
Moreover, the geometry of pure zigzag edges are found to be metastable.
It was shown that the planar reconstruction of the zigzag edge to reczag type is 
possible because it lowers the edge energy \cite{koskinen2008}.
The electronic structure of graphene with reczag edges 
differs significantly from that with zigzag edges, namely by 
the appearance of the degeneracy of conduction and valence bands
for the momentum range\cite{wassmann2008}, $k < 2\pi /3$, and their dispersion for $k > 2\pi /3$ 
(the zigzag edges imposes the flat degenerate conduction and valence bands 
within the interval \cite{nakada1996}, $2\pi/3 \le \mid k\mid \le \pi$).
Coexistence of the zigzag and reczag edges in graphene samples
has been confirmed experimentally \cite{koskinen2009}.

In fabrication of graphene-based devices 
the control of the shape and quality of the edges is clearly an important issue.
With the standard approaches developed to create graphene nanoribbons,
such as lithographic patterning \cite{lit}, 
chemical vapor deposition method \cite{vapour} and chemical 
sonication \cite{sonic}, the control of the 
nanoribbon size and quality of the edges is poor.
Most recent achievement in 
fabrication of graphene nanoribbons is by unzipping 
the carbon nanotubes \cite{nature1,nature2}. 
Unzipping of multiwalled or singlewalled carbon
nanotubes by plasma etching provides smooth edges and 
a small range of the nanoribbon width \cite{nature1} (10-20 nm).
The second unzipping method - the solution-based oxidative 
process allows to perform the longitudinal cut of the nanotubes 
and to obtain the nanoscale structures characterized 
by the predominantly straight linear edges \cite{nature2}.
However, because of the applied solution-based oxidative method,
the zigzag edges of graphene are oxidized and so the 
surface of graphene can be partially oxidized.
Annealing or chemical reduction of samples 
in the N$_2$H$_4$ environment reduces 
the number of oxygen-containing functionals 
at the surface and edges of graphene and, indeed, a 
significant increase in conductivity of graphene has been obtained \cite{nature2}.

From our point of view, an important improvement 
of oxidation of the zigzag edges of graphene is the prevention of
reconstruction of the zigzag edges. 
However, an extensive investigation of 
the electronic properties of graphene with oxidized zigzag edges
is not yet available in the literature. 
Therefore, the present study is devoted to this particular issue and 
the possibility of doping nanoscale graphene 
through adsorption of water and gas molecules.
The spin-polarized density functional theory with semilocal gradient 
corrected functional (UB3LYP/6-31G \cite{b3lyp}) in the Jaguar 6.5 program \cite{jaguar}
has been applied in our work. The impact of the van der Waals interactions 
is not included within the DFT that leads to underestimation 
of the adsorption energy \cite{ort}, but it should not undermine 
the reliability of DFT
for investigation of the influence of adsorption 
on the electronic properties of graphene. 

\section{\label{man} Unique properties of graphene with oxidized zigzag edges}
\begin{figure*}
\includegraphics[scale=0.85]{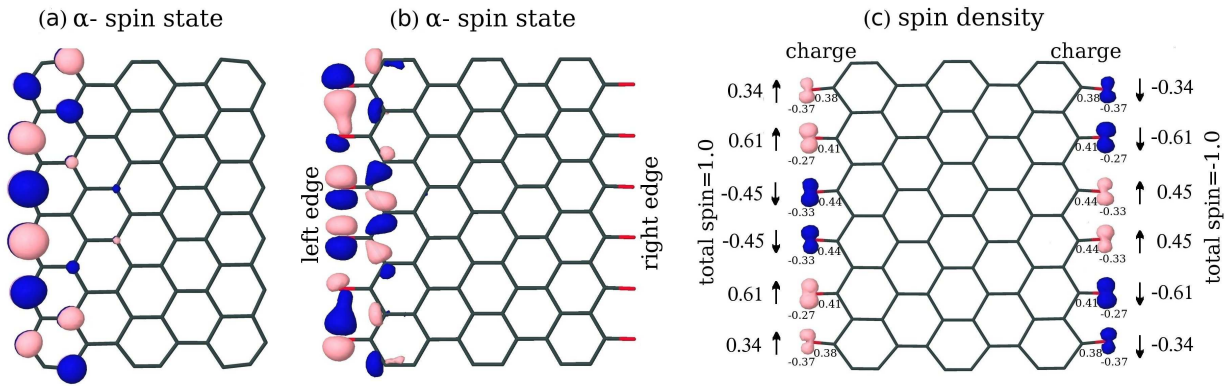}
\caption{\label{fig:fig1} (color online) 
Electron and spin density distribution in a pure graphene flake and 
graphene with oxidized zigzag edges. 
The electron and spin densities are plotted for isovalues of $\pm0.03$ e/\AA$^{3}$.
(a) The electron density distribution in the pristine graphene for a state of C$_{2v}$ symmetry
($\alpha$-spin for HOMO).
(b) The electron density distribution in graphene with oxidized zigzag edges for a state
of C$_{1}$ symmetry ($\alpha$-spin for HOMO-9).
(c) The spin density distribution in graphene with oxidized zigzag edges for a state
of C$_{1}$ symmetry. Localization of spins 
on the oxygen atoms obtained within the NBO analysis
and the natural charge for the oxygen and carbon atoms within the
C-O bonds are also indicated.}
\end{figure*}

For graphene of finite size with pristine edges, 
the effect of localization of the electron density at the zigzag edges 
is a result of specific boundary conditions for the wavefunctions. 
In particular, because all border carbon atoms across a single zigzag edge belong 
to the same sublattice, the wavefunctions at the zigzag edge vanish on a single sublattice,
while at the armchair edges on both sublattices. 
The generated localized states at the zigzag edges 
are responsible for formation of the conduction and valence bands near the Dirac points.
Depending on the spin ordering of the localized states
between the zigzag edges, graphene of finite size can 
be gapless or has a gap.
If spins of the localized electrons are antiferromagnetically 
ordered between the opposite zigzag edges for which 
the border carbon atoms belong to
different sublattices, the sublattice symmetry is preserved.
In the case of an absence of the confinement effect, 
graphene with preserved sublattice symmetry is expected to be gapless.
The ferromagnetic ordering between the zigzag edges leads to
breaking of the sublattice symmetry and opening of a gap.
In the case of broken sublattice symmetry 
(for nanoscale graphene the sublattice symmetry is 
broken already for C$_{2v}$ symmetry when the mirror plane of symmetry is
perpendicular to the zigzag edges \cite{julia,julia1}), 
the $\alpha$-spin state is localized on one zigzag edge, while 
the $\beta$-spin state on the opposite zigzag edge. 
We presented an example of the distribution of 
the electron density of the localized $\alpha$-spin state 
for the highest occupied molecular orbital (HOMO) 
in Fig.~\ref{fig:fig1} (a). A similar distribution but 
with localization of the electron density on 
the opposite zigzag edge is observed
for the lowest unoccupied molecular orbital (LUMO).
The maximum of the electron density of 
these states is located on the border carbon atoms and, 
as a result, the spin density is localized there as well 
(see Fig.2 in Ref.\cite{julia}). 

Oxidation of the zigzag edges makes a difference 
in the electron density distribution of the localized states 
because of alteration of the boundary conditions for the wavefunctions.
The electron density is mostly localized on the sides of the polar C-O bonds,
as it is presented in Fig.~\ref{fig:fig1} (b). 
For the structure shown in Fig.~\ref{fig:fig1} (b), 
the electron density for the HOMO and LUMO are delocalized over the 
whole graphene structure, while the localized states 
are shifted deeper into the bands from the gap edges, i.e.
to the HOMO-9 and HOMO-10 in the valence band 
and to the LUMO+1 and LUMO+2 in the conduction band.
However, an increase in the size of the structure leads to
backward shifting of these states closer to the band gap. 
Therefore, when we increased the width of 
nanoscale graphene by adding two rows of the carbon hexagons 
along the single zigzag edge, the localized states 
in the valence band has already been shifted
to HOMO-5 and HOMO-6 orbitals.

The spin density for graphene with oxidized edges is 
mostly localized on the oxygen atoms, as it is presented
for nanoscale graphene in the C$_{1}$ symmetry state in Fig.~\ref{fig:fig1} (c).
Ordering of spins of the localized electrons along the zigzag edges and between
opposite edges is defined by the applied symmetry. For the 
$D_{2h}$ symmetry, the spin ordering is ferromagnetic along the edges and between 
the edges, while for the $C_{2v}$ symmetry 
(when the symmetry plane is perpendicular to the zigzag edges) 
and for the $C_{1}$ symmetry, there are mixed spin ordering along 
the zigzag edges and antiferromagnetic ordering 
between the zigzag edges.
The total energy of the system decreases with lowering of the 
symmetry of graphene and the $C_{1}$ symmetry state
is characterized by the lower energy.
For the state of the $C_{1}$ symmetry, 
the total spin at each zigzag edge is nonzero, particularly 
the total spin equals +1.0 for one edge and -1.0 for another edge.
Because the spin density is mostly localized on the oxygen atoms 
within the C-O bonds, it is negligible for all carbon atoms in graphene.
Symmetry also has influence on the size of the band gap. 
For the size of graphene presented in Fig. 1 (b), 
the gaps are $\Delta_{D_{2h}}$=0.18 eV, $\Delta_{C_{2v}}$=1.0 eV 
and $\Delta_{C_{1}}$=1.33 eV.
We consider that graphene obtained by nanotube unzipping 
would have highly tensile structure for which the 
high symmetry state would be unreachable, and therefore,
we have used the C$_{1}$ symmetry for all our calculations. 

We have performed the natural bond analysis \cite{nbo} (NBO) for nanoscale 
graphene with the oxidized edges
and defined the natural charges for the carbon and oxygen atoms within 
the C-O bonds, whose magnitudes are indicated in Fig.~\ref{fig:fig1} (c). 
The C-O is the polar bond for which the decentralization
of the electron clouds between the carbon and oxygen atoms 
leaves the oxygen atoms negatively charged, 
while the carbon atoms are positively charged.
As a result, the electro-negative oxygen atoms within 
the polar C-O bonds are more chemically active 
than on the non-polar surface of graphene.
It was found that for the C-O bonds the 
natural charge distribution (see Fig.~\ref{fig:fig1} (c))
and the length of the bonds deviate along the zigzag edges (from 1.26 to 1.31 \AA). 
However, despite the dispersion of the bond length 
the variation of the dipole moment among the C-O bonds 
is consistent with the deviation of the natural charge distribution, 
i.e. for the bonds characterized by a larger charge difference 
between the oxygen and carbon atoms, 
the dipole moment is slightly larger as well. 
Therefore, interaction of the oxygen atoms with the adsorbed molecules 
would also slightly deviate along the edges. 

\section{Adsorption of $\mathrm{H_{2}O}$ and $\mathrm{NH_3}$ 
molecules}

The most popular and widely investigated 
method of doping graphene is by adsorption of the 
gas molecules which is claimed to be 
controllable by the nature and 
concentration of the adsorbates 
\cite{schedin2007,wehling2008,dongfu2009,voggu2008,gierz2008,bost1,ohta-sci313}.
However, according to the theoretical predictions, interaction between the
surface of graphene and most of the molecules adsorbed on graphene 
should be low \cite{ort}
and doping of ideal pristine graphene (sp$^2$ network) 
should not be efficient. 
Moreover, because of weak interaction of the adsorbates 
with the ideal graphene surface 
the type of doping and its efficiency is defined
by the orientation of the adsorbed molecules on the surface
\cite{julia,leenaerts,leenaerts1,huang}.
Therefore, doping observed in experiments 
\cite{schedin2007,wehling2008,dongfu2009,voggu2008,gierz2008,bost1,ohta-sci313} 
is most probably the result of lattice defect in graphene \cite{defects}. 
In this respect we would like to emphasize two important advantages of 
oxidation of the zigzag edges of graphene: 
potentials to achieve a controllable doping (because 
oxidation will enhance interaction of the edges 
with the surrounding environment)
and stabilization of the edge geometry.
The oxidized zigzag edges attract polar molecules 
from the environment more strongly than the graphene surface
due to polarity of the C-O bonds. 
The charge exchange between the C-O bonds and the adsorbates would 
contribute to the alteration of 
the electronic balance in graphene leading to its doping. 
Therefore, adsorption would influence the graphene conductivity 
and offers an unique opportunity for its application in
graphene-based gas sensors \cite{schedin2007,wehling2008}. 

\begin{figure}
\includegraphics[scale=0.45]{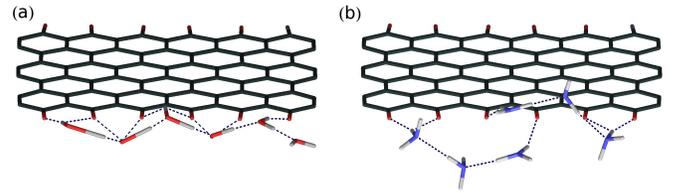}
\caption{\label{fig:fig2} (color online) Structure of nanoscale graphene
with oxidized edges where the adsorbed molecules are attached
to the zigzag edge by hydrogen bonds. Several adsorbed molecules form 
a cluster due to the hydrogen bonding with zigzag edge of the oxidized graphene and 
between each other. 
(a) adsorbate is the $\mathrm{H_{2}O}$ molecule. (b) adsorbate is the $\mathrm{NH_3}$ molecule.}
\end{figure}

The $\mathrm{H_{2}O}$ or $\mathrm{NH_3}$ molecules 
have been chosen for our investigation 
because of their polarity and the presence of more than 
two hydrogen atoms in their structure. 
Adsorption of $\mathrm{H_{2}O}$ or $\mathrm{NH_3}$ molecules 
on graphene with oxidized edges has been performed for 
primary optimized geometry of graphene.
After adsorption an additional optimization procedure 
has been applied. During this optimization the
positions of the carbon atoms belonging
to graphene are kept fixed, 
while the oxygen atoms within the C-O bonds
and the adsorbed molecules were relaxed.
The oxygen atoms within the C-O bonds
make hydrogen bonds with the adsorbed molecules
due to strong polarity of the C-O bonds.
If adsorbed molecules have in their structure 
several hydrogen atoms and the electro-negative atoms 
possess the lone pair (oxygen, nitrogen), 
they have the ability to bind to each other through the hydrogen bonds,
thereby forming a cluster. 
However, the hydrogen bonding with graphene leads to aligning 
of the cluster structure along the zigzag edge, as presented 
in Fig.~\ref{fig:fig2} (a) and  Fig.~\ref{fig:fig2} (b)
for the $\mathrm{H_{2}O}$ and $\mathrm{NH_3}$ molecules, respectively.
We have estimated the binding energy required to dissociate
graphene from the adsorbed molecules. It was found that
the binding energy grows almost linearly with 
increasing number of the adsorbed molecules,
but have a slight tendency for saturation when the number 
of molecules is larger than eight.

The highest symmetry applicable for graphene 
with adsorbates is the $C_{1}$ symmetry 
because the charge exchange with the adsorbates generates 
an imbalance in the electron distribution 
between the two sublattices in graphene and, therefore, 
breaking the sublattice symmetry. 
The size of the bandgap after adsorption
is similar to the value of that for pristine graphene 
in the $C_{1}$ symmetry state ($\Delta_{C_{1}}$=1.33 eV). 
However, if the HOMO or LUMO are localized 
on the adsorbed molecules, the size of the band gap is altered. 
Interaction of graphene with the adsorbed molecules induces a 
change in ordering of the orbitals, particularly 
the states localized on the zigzag edges now belong to 
HOMO (HOMO-1) and LUMO (LUMO+1) orbitals while
without adsorbates the localized states are shifted 
deeper into the conduction and valence bands 
(see discussion in Sec. \ref{man}).

\begin{table}
\caption{\label{tab:table1} Influence of adsorption 
of $\mathrm{H_{2}O}$ or $\mathrm{NH_{3}}$ molecules 
on spin and charge distributions in 
graphene with oxidized edges.
All data are obtained within the NBO analysis.
$Q_{\mathrm{H_{2}O}}$ and $Q_{\mathrm{NH_{3}}}$ 
are the charge accumulated on the adsorbed molecules, 
$S_{\mathrm{H_{2}O}}$ and $S_{\mathrm{NH_{3}}}$ are
the total spin on the cluster formed by the adsorbed molecules, 
$S_{\mathrm{O}}$ is the sum of spin located on the oxygen atoms 
belonging to the polar C-O bonds of graphene, 
$Q_{tr}$ is the total charge transfer from graphene to the adsorbates.
"Left" and "Right" refer to  
to the left and right zigzag edges as shown in Fig.~\ref{fig:fig1} (b).}
\begin{ruledtabular}
\begin{tabular}{c|r|r|r|r|r|r|r}
\multicolumn{7}{c}{$\mathrm{H_{2}O}$} \\
\hline 
$N_{\mathrm{H_{2}O}}$ 
& \multicolumn{2}{c|}{$Q_{\mathrm{H_{2}O}}$, \=e} 
& \multicolumn{2}{c|}{$S_{\mathrm{H_{2}O}}$}
& \multicolumn{2}{c|}{$S_{\mathrm{O}}$}
& $Q_{tr}$, \=e \\
& Left
& Right 
& Left
& Right
& Left 
& Right
& \\
\hline 
1 & \multicolumn{2}{l|}{0.028} & \multicolumn{2}{l|}{0.00} & 0.00 & 0.00 &  0.006 \\
2 & \multicolumn{2}{l|}{0.317} & \multicolumn{2}{l|}{-0.46} & -0.47 & 1.01 & 0.028 \\
3 & \multicolumn{2}{l|}{0.841} & \multicolumn{2}{l|}{0.48} & -1.59 & 1.01 & 0.084 \\
4 & \multicolumn{2}{l|}{0.856} & \multicolumn{2}{l|}{-1.02} & 2.01 & -1.01 & 0.068 \\
5 & \multicolumn{2}{l|}{0.869} & \multicolumn{2}{l|}{1.02} & -2.01 & 1.00 & 0.096 \\
6 & \multicolumn{2}{l|}{0.889} & \multicolumn{2}{l|}{1.00} & 0.01 & -1.00 & 0.073 \\
\hline 
7  & 0.882 & -0.006 & 1.00 & -0.01 & 0.01 & -1.00 & 0.096 \\
8  & 0.890 & 0.053 & 1.00 & -0.07 & 0.01 & -0.93 & 0.098 \\
9  & 0.881 & 0.348 & -1.00& 0.40 & 0.04 & 0.24 & 0.132 \\
10 & 0.888 & 0.854 & 1.01 & 0.99 & -1.97 & 0.01 & 0.161 \\
11 & 0.888 & 0.864 & 1.01 & 1.01 & 0.01 & -1.97 & 0.183 \\
12 & 0.889 & 0.853 & 1.00 & 1.01 & 0.01 & -1.95 & 0.191 \\
\hline
\multicolumn{7}{c}{$\mathrm{NH_{3}}$} \\
\hline
$N_{\mathrm{NH_{3}}}$ 
& \multicolumn{2}{c|}{$Q_{\mathrm{NH_{3}}}$} 
& \multicolumn{2}{c|}{$S_{\mathrm{NH_{3}}}$}
& \multicolumn{2}{c|}{$S_{\mathrm{O}}$}
& $Q_{tr}$ \\
& Left & Right & Left & Right & Left & Right
& \\
\hline 
1 & \multicolumn{2}{l|}{0.247} & \multicolumn{2}{l|}{-0.26} & -0.74 & 1.00 & -0.018 \\
2 & \multicolumn{2}{l|}{0.906} & \multicolumn{2}{l|}{1.00} & -0.01 & -1.03 & 0.035 \\
3 & \multicolumn{2}{l|}{0.922} & \multicolumn{2}{l|}{1.00} & 0.01 & -1.00 & 0.057 \\
4 & \multicolumn{2}{l|}{0.925} & \multicolumn{2}{l|}{1.00} & 0.01 & -1.00 & 0.052 \\
5 & \multicolumn{2}{l|}{0.933} & \multicolumn{2}{l|}{1.00} & -2.00 & 1.00 & 0.043 \\
6 & \multicolumn{2}{l|}{0.926} & \multicolumn{2}{l|}{1.00} & 0.01 & -1.01 & 0.048 \\
\hline 
7 & 0.925& 0.271 & -1.00 & 0.28 & -0.01 & 0.72 & 0.054 \\
8 & 0.921 & 0.903 & -1.00 & 1.00 & 0.01 & -0.01 & 0.127 \\
9 & 0.921 & 0.922 & -1.00 & 1.00 & -0.01 & 0.01 & 0.109 \\
10 & 0.924 & 0.921 & 1.00 & -1.00 & -0.01 & 0.01 & 0.100 \\
11 & 0.924 & 0.928 & -1.00 & -1.00 & 1.99 & -0.02 & 0.094 \\
12 & 0.924 & 0.924 & -1.00 & 1.00 & -0.01 & 0.01 & 0.097 \\
\hline 
\end{tabular}
\end{ruledtabular}
\end{table}

Formation of hydrogen bonds between the adsorbed molecules and 
the oxygen atoms belonging to the C-O bonds of graphene leads to 
decentralization of the electron clouds between 
the electro-negative oxygen from the C-O bonds and the electro-positive 
hydrogens belonging to the adsorbed molecules.
Therefore, the interaction of the adsorbates and graphene triggers 
a charge redistribution between them.
Accumulation of the natural charge on the adsorbed molecules 
determined within the NBO analysis is presented 
for $\mathrm{H_{2}O}$ ($Q_{\mathrm{H_{2}O}}$) 
and $\mathrm{NH_3}$ ($Q_{\mathrm{NH_{3}}}$) 
molecules in Table ~\ref{tab:table1}.
The charge exchange between graphene and the adsorbates 
has been also defined within the NBO procedure as $Q_{tr}=Q_{DA}-Q_{AD}$,
where $Q_{DA}$ is the charge transfer from graphene to the adsorbates and 
$Q_{AD}$ is the charge transfer from the adsorbates to graphene. 
The $Q_{DA}$ and $Q_{AD}$ parts have been calculated 
as the sum of the $\Omega_i\rightarrow\Omega^*_j$ 
charge transfers between the donor orbital $\Omega_i$ and the
acceptor orbital $\Omega^*_j$ defined within the NBO procedure as:
\begin{equation}
Q_{DA}=\sum_{i,j}Q_{i,j}=\sum_{i,j}q_i F_{i,j}^2/(\epsilon_i-\epsilon_j)^2
\label{eq:one}
\end{equation}
where $q_i$ is the donor orbital occupancy, 
$\epsilon_i,\epsilon_j$ are the orbital energies, 
$F_{i,j}$ is the off-diagonal element. The charge occupancy transfer is considered for 
stabilizing orbital interactions, i.e. when the second order interaction energy $\Delta
E_{i,j}=-2F_{i,j}^2/(\epsilon_i-\epsilon_j)$ has a positive sign. 
The calculated results for the charge exchange ($Q_{tr}$) between 
graphene and the adsorbates ($\mathrm{H_{2}O}$ or $\mathrm{NH_3}$ molecules)
are also presented in Table ~\ref{tab:table1}.
The total spin of the states localized on the oxygen atoms 
within the C-O bonds ($S_{\mathrm{O}}$) and on the adsorbed molecules 
($S_{\mathrm{H_{2}O}}$ or $S_{\mathrm{NH_{3}}}$)
are also indicated in this table separately for
left and right zigzag edges of graphene (see notations for the edges in 
Fig.~\ref{fig:fig1} (b)).

\begin{figure}
\includegraphics[scale=0.35]{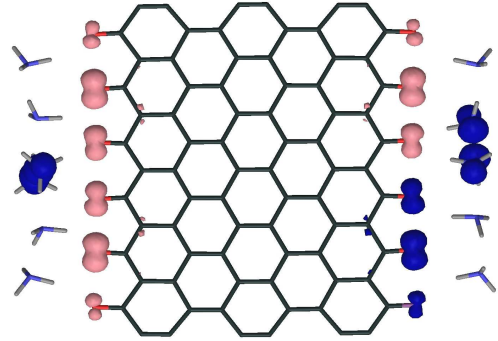}
\caption{\label{fig:fig3} (color online) The spin distribution in 
graphene with eleven adsorbed NH$_3$ molecules 
plotted for isovalues of $\pm0.01$ e/\AA$^{3}$.
The left edge has a total spin 2.0, 
while for the right edge the total spin is zero. 
The adsorbed molecules maintain spin of -1.0 from 
each side of graphene, 
however the balance of the total spin between two sides 
of graphene with the adsorbates is maintained,
i.e., -1.0 for one side and 1.0 for other side.}
\end{figure}

According to our studies, 
both graphene and the cluster formed by the adsorbed molecules 
are spin-polarized in most cases. 
For spin-polarized graphene,
each zigzag edge is characterized by the non-zero total spin
whose sign is different for the opposite edges.
The sign and the total spin associated with the adsorbed molecules 
varies with the number of molecules.
However, the balance of the spin density between 
the left and right sides of the whole system which includes 
graphene with adsorbed molecules, is preserved
(except for the case when a single water molecule is adsorbed). 
Thus, taking into account the 
spin localized on the zigzag edge and the adsorbed molecules 
(see the magnitude of $S_{\mathrm{NH_{3}}}+S_{\mathrm{O}}$ and 
$S_{\mathrm{H_{2}O}}+S_{\mathrm{O}}$ in Table ~\ref{tab:table1})
the total spin +1.0 for one side of the system 
and -1.0 for other side are retained.
The spin distribution for the oxidized graphene with eleven
NH$_3$ molecules bound to the edges by the hydrogen bonds 
is presented in Fig.~\ref{fig:fig3}. Only one or two molecules
within the cluster have the non-zero total spin 
and are spin-polarized, while the rest of the molecules are unpolarized. 
For adsorbed molecules characterized by zero spin density 
the magnitude of the natural charge is almost zero ($\sim$ 0.05 \=e), 
but for a case of non-zero spin density the charge is non-zero 
($\sim$ 0.5 \=e for each of the spin-polarized molecule).
The dependence of spin-polarization and resulting nonzero
magnetic moment on the number and location of the adsorbed molecules
can be useful for developing spin-valve devices \cite{cho}.

The charge transfer from pristine graphene to water
is not expected to be efficient 
and was reported to be in the range \cite{leenaerts},
-0.02 $\div$ 0.01 \=e depending on the
position and orientation of water, 
while according to our calculations
performed for the single water molecule within the NBO analysis,  
it is $\sim$ 0.001 \=e (for details of interaction of 
water with pristine graphene see \cite{julia}).
The charge exchange between graphene with oxidized edges 
and the adsorbates is quite efficient and
increases with the number of adsorbed molecules (see $Q_{tr}$ in Table ~\ref{tab:table1}).
For adsorption of H$_2$O molecules, the charge donation from graphene to 
adsorbates increases gradually.
However for a single H$_2$O molecule adsorbed at the zigzag edge,
the charge exchange is insignificant because of the
location of the molecule relative to the edges. 
A single H$_2$O molecule interacts with graphene 
by both of its hydrogens which are directed to two nearest oxygen atoms
at the zigzag edge. The resulting distance between the H$_2$O molecule and 
the edges is large (the distance between O-H and O-C is $\sim$ 2.2 \AA) 
and the H$_2$O molecule lies parallel to the graphene plane. Therefore,
the overlapping of orbitals is weak.
For two or more adsorbed molecules, one hydrogen of the H$_2$O molecule 
makes a hydrogen bond with the other H$_2$O molecule, while the second 
hydrogen bonds to the oxygen atom within the C-O bond of graphene. 
The cluster that formed is placed close to the zigzag edge 
(the hydrogen bond of O-H $\cdot\cdot\cdot$ O-C is $\sim$ 1.2 \AA)
and water molecules are located not exactly in the plane 
of graphene but slightly below and above, hence providing 
an effective overlapping of orbitals between 
the interacting species (see Fig.~\ref{fig:fig2} (a)). 

\begin{figure}
\includegraphics[scale=0.45]{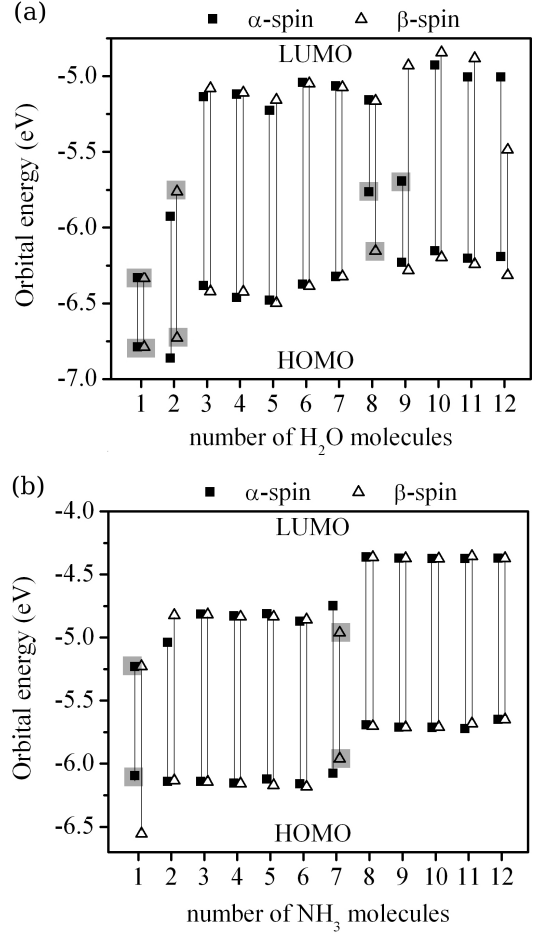}
\caption{\label{fig:fig4} Effect of adsorption on the 
orbital energies of the HOMO and LUMO. 
The results are presented for both 
the $\alpha$- and $\beta$-spin states. 
(a) adsorbate is the H$_2$O molecule. 
(b) adsorbate is the NH$_3$ molecule.
The shaded areas are used to mark orbitals 
which are localized on the adsorbates.}
\end{figure}

For the NH$_3$ adsorbates, a similar behavior 
is obtained for up to three NH$_3$ molecules. 
The most efficient charge exchange occurs 
when two or three molecules are placed along the
zigzag edge, while a further increase 
of their number suppresses the charge transfer.
The effect is related to clustering of the NH$_3$ molecules 
along the zigzag edge. 
Two or three NH$_3$ molecules adsorbed along the zigzag edge 
are located slightly below and above the graphene plane
somewhat similar to the water molecules in Fig.~\ref{fig:fig2} (a)
(the hydrogen bond of N-H $\cdot\cdot\cdot$ O-C is $\sim$ 2.0 \AA)
generating an efficient charge exchange with graphene due to 
strong overlapping of their orbitals.
However, adsorbates move apart from the plane of graphene 
when more than three molecules are added 
(as presented in Fig.~\ref{fig:fig2} (b)) and the length of 
hydrogen bonds for some molecules increases 
(the hydrogen bond of N-H $\cdot\cdot\cdot$ O-C is $\sim$ 2.5 \AA).
Thus, in the cluster containing more than three NH$_3$ molecules,
they are located significantly out of the plane 
of graphene that lead to 
diminishing efficiency of the charge exchange 
of each adsorbed molecule with graphene.

The charge exchange between graphene and the adsorbed molecules
shifts the conduction and valence bands of graphene in the energy scale. 
In Fig.~\ref{fig:fig4}, we have plotted 
the orbital energies of HOMO and LUMO as a function of the 
number of adsorbed molecules . The HOMO-LUMO gap 
is almost unchanged except for the cases when the orbitals are localized on
the adsorbed molecules as indicated in Fig.~\ref{fig:fig4} by the shaded areas.
The orbital energies are shifted when the number of molecules increases
indicating the $p$-doping of graphene. Degradation of the bands 
is clearly consistent with the alteration 
of the charge transfer from graphene to the adsorbates 
(see $Q_{tr}$ in Table ~\ref{tab:table1}).
Thus for the adsorption of water, the shift 
of the orbital energy gradually increases when the number of 
adsorbed molecules grows, thereby providing an 
opportunity to control the doping of graphene 
through adsorption of the polar molecules. 

\section{Conclusion}

We have investigated the alteration of the 
electronic properties and spin distribution 
induced by adsorption of the polar molecules in 
nanoscale graphene with oxidized zigzag edges.
The polar molecules interact 
with oxygen atoms belonging to the polar C-O bonds 
at the edges of graphene. When the adsorbed molecule 
contains hydrogen atoms they interact 
with the C-O bonds of graphene through generation of 
the hydrogen bonds (H $\cdot\cdot\cdot$ O-C).
Several adsorbed $\mathrm{H_{2}O}$ or $\mathrm{NH_3}$ molecules 
tends to create a cluster 
due to the strong intermolecular interaction between them,
but the interaction with graphene 
aligns the cluster structure along the zigzag edge.
Graphene with oxidized edges tends to donate 
a charge to the adsorbates and its efficiency is defined 
by the intermolecular distance and 
by the location of the adsorbed molecules relative to the 
plane of graphene. 
In the case of H$_2$O adsorption, 
the charge transfer efficiency and the corresponding shift of the 
HOMO and LUMO bands as a result of $p$-doping of graphene 
increase gradually with the number of molecules. 
For the NH$_3$ molecules the maximum efficiency 
of doping occurs when the number of adsorbed molecules along the zigzag edge 
is limited by two or three as a result of their 
closer location to the plane of graphene than when more molecules 
are adsorbed.
Therefore, we can conclude that when the graphene surface 
or the edges are oxidized, the adsorption of the polar molecules 
is the most viable option for $p$-doping of graphene.

\section{Acknowledgments}
The work was supported by the Canada Research Chairs
Program and the NSERC Discovery Grant.\\

\end{document}